\documentclass[pra,superscriptaddress,showpacs,twocolumn]{revtex4}
\usepackage{graphicx}          
\begin{document}
\title{Joint measurements and Bell inequalities}
\author{Wonmin Son}
\affiliation{School of Mathematics and Physics, Queen's University,
Belfast BT7 1NN, United Kingdom}
\author{Erika Andersson}
\affiliation{Department of Physics, University of Strathclyde,
Glasgow G4 0NG, UK}
\author{Stephen M. Barnett}
\affiliation{Department of Physics, University of Strathclyde,
Glasgow G4 0NG, UK}
\author{M. S. Kim} \affiliation{School of
Mathematics and Physics, Queen's University, Belfast BT7 1NN, United
Kingdom}
\date{September 20, 2005}

\begin{abstract}
Joint quantum measurements of non-commuting observables are possible, if one accepts an increase in the measured variances. A necessary condition for a joint measurement to be possible is that a joint probability distribution exists for the measurement. This fact suggests that there may be a link with Bell inequalities, as these will be satisfied if and only if a joint probability distribution for all involved observables exists. We investigate the connections between Bell inequalities and conditions  for joint quantum measurements to be possible. Mermin's inequality for the three-particle Greenberger-Horne-Zeilinger state turns out to be equivalent to the condition for a joint measurement on two out of the three quantum systems to exist. Gisin's Bell inequality for three co-planar measurement directions, meanwhile, is shown to be less strict than the condition for the corresponding joint measurement.
\end{abstract}


\maketitle

\section{Introduction}
A joint quantum measurement means that by performing {\it one} measurement on a {\it single} quantum system, we are able to produce a result for {\it each} of two observables \cite{artkel, artgood, stig}. The two measured observables do not have to commute with each other. A trivial example of such a measurement would be to measure one of the observables, and randomly guess a result for the other observable. It is, however, also possible to perform joint measurements where the element of guessing is distributed more equally between the two non-commuting observables.

A necessary condition for the existence of a joint quantum measurement is that a joint probability distribution exists for the measurement. This probability distribution has to give the correct marginal distributions, and has to exist for any measured quantum state. For a joint quantum measurement, this will force the marginal probability distributions to have greater variances than if the observables were measured alone \cite{artkel,artgood}. For example, the Wigner function of two complementary observables may be negative for some quantum states, and is therefore not a true probability distribution. In order to obtain a positive joint probability distribution for the two observables, we have to consider the $Q$ function \cite{stig, walker}. The marginal distributions of the $Q$ function have greater spread than those of the Wigner
function \cite{barnett}.

Bell's great achievement was to derive an inequality which had to be satisfied in any local realistic theory \cite{bell}. The inequality gives a bound on the measurement correlations for a bipartite system. The requirement that the Bell inequalities are satisfied is equivalent to the existence of a joint probability
distribution for the observables involved \cite{fine}. A necessary condition for a joint quantum mechanical measurement of these observables is therefore that the Bell inequalities hold. This condition may, however, not be sufficient.

In an earlier paper \cite{jointspin}, the connection between the restrictions placed on joint quantum measurements and the Clauser-Horne-Shimony-Holt (CHSH) inequality \cite{chsh} were investigated. In the CHSH setup, one considers a pair of spin 1/2 particles; on each one, one makes a choice of measuring a spin along one of two possible directions. It turns out that if we instead make a joint measurement on {\it one} of the two spin 1/2 quantum systems involved in the setup, then the CHSH inequality is equivalent to the condition for this joint measurement to be possible. Evidently also if the measurements on both particles are joint measurements, then the Bell inequalities will be satisfied \cite{demuynck}.
In this paper, we consider the connection between joint quantum measurements and Mermin's inequality \cite{mermin} for the three-particle Greenberger-Horne-Zeilinger (GHZ) state \cite{greenberger}. We show that if joint measurements are made on two of the three quantum systems, then the resulting condition for the existence of such a measurement is equivalent to Mermin's inequality.

As already mentioned above, Bell inequalities being satisfied by the measurement statistics is not necessarily a sufficient condition for a joint measurement to exist. As an example of this, we consider Gisin's inequality for three co-planar measurement directions \cite{gisinbell}. We shall find that the
requirement for a joint quantum measurement of a spin along the considered three directions to be possible is {\it stronger} than the corresponding Bell inequality.

The paper is organized as follows. In Section \ref{jointsec}, we give a brief account of joint quantum measurements, concentrating on joint measurements of two components of spin 1/2 \cite{busch},
and in Section \ref{chshsec} we remind the reader of the CHSH inequality. In Section \ref{ghzsec}, we proceed to discuss Mermin's inequality for GHZ states. Here we show that making joint quantum measurements of the different spin components means that Mermin's inequality must be satisfied. In Section \ref{gisinsec}, we give examples of when the restrictions for the existence of joint quantum measurements are strictly stronger than the conditions placed by local realism. We conclude
with a discussion in Section \ref{conclsec}.

\section{Joint measurements of non-commuting observables}
\label{jointsec} 

Quantum measurements are usually described in standard quantum mechanics textbooks as projective measurements. A measurement of an observable $\hat{A}$ is a projection in the eigenbasis of $\hat{A}$, and the measurement outcomes are the corresponding eigenvalues. Restricting ourselves to projective quantum measurements means that it is not possible to measure two non-commuting observables jointly at the same time, as they do not have a common set of eigenstates.

Projective measurements, however, are too restrictive. The framework of generalized quantum measurements, referred to as probability operator measurements (POMs) or positive operator valued
measures (POVMs) \cite{hel, nch}, allows us to express conveniently exactly what measurements it is possible to realize within quantum mechanics. Allowing for detector inefficiencies, for example, usually means that the measurement is not truly projective, but has to be described as a generalized measurement. Such a measurement is, just like a projective measurement, described by a set of measurement operators, one for each measurement outcome. There is no condition, however, saying that the measurement operators have to be projectors onto eigenstates. The measurement operators must have only positive or zero eigenvalues, thus guaranteeing that the probability for each outcome will be positive or zero for any measured quantum state. The operators should also sum to the identity operator, corresponding to the fact that the sum of all probabilities for all possible experimental outcomes is 1.

It is possible to measure non-commuting observables jointly, if one allows an increase in the variances of the jointly measured observables.  Pioneering work on joint quantum measurements was done in \cite{artkel, artgood}, and an elucidating account is found in \cite{stig}.  Such measurements are most conveniently described as generalized quantum measurements.

A frequently used condition for when a joint measurement is a ``good" measurement of two observables $\hat{A}$ and $\hat{B}$ is that the expectation values of the jointly measured observables should be proportional to the expectation values of the observables when measured alone. We write
\begin{equation}
\overline{A_J} = \alpha\langle \hat{A}\rangle ~~~\mbox{and}~~~
\overline{B_J} = \beta\langle\hat{B}\rangle,
\end{equation}
where $\alpha$ and $\beta$ are positive constants and where $\overline{A_J}$ and $\overline{B_J}$ are expectation values of the jointly measured observables.
(Here we do not use the notation $\langle \cdot\rangle$ for an expectation value, as $A_J$ and $B_J$ are not observables in the usual sense.) A sharp measurement of spin 1/2 particle along a
direction $\bf a$ corresponds to a projective measurement of the observable $\hat{A}=\bf a\cdot{\vec{\sigma}}$, where the Pauli spin operator $\vec{\sigma}=(\hat\sigma_x, \hat\sigma_y,
\hat\sigma_z)$. Busch and co-workers \cite{busch} found that the condition for a joint measurement of two spin 1/2 components, parallel to the unit vectors $\bf a$ and $\bf b$, to be possible, is
\begin{equation}
\label{conjoint} {|\alpha\bf a +\beta b| + |\alpha a -\beta b|}
\leq 2.
\end{equation}
Unless $\bf a$ and $\bf b$ are parallel, this condition will restrict the values of $\alpha$ and $\beta$ to be strictly less than one. As the possible values for the measurement outcomes are $\pm 1$, this means that the variances
\begin{eqnarray}
\label{simvar}
(\Delta{{A}_J})^2&=&\overline{{A}_J^2} - \overline{{A}_J}^2
= 1- \alpha^2\langle{\hat{A}}\rangle^2\nonumber\\
(\Delta{{B}_J})^2&=&\overline{{B}_J^{2}} - \overline{{B}_J}^2 =
1- \beta^2\langle{\hat{B}}\rangle^2
\end{eqnarray}
must increase. An earlier work \cite{jointspin} investigated the connection between the joint measurements of a spin along two directions and the CHSH Bell inequalities, as well as the increase in uncertainty for the jointly measured observables. In the following Section, we review the derivation
of the CHSH inequality and its connection with joint measurements.

\section{The CHSH inequality}
\label{chshsec}

In the CHSH setup, we consider two spin 1/2  particles. Spin is measured along either direction ${\bf a}_1$ or ${\bf b}_1$ on system 1, and along either direction ${\bf a}_2$ or ${\bf b}_2$ on system 2 \cite{chsh}. Let us denote the measurement outcomes by $a_1, b_1, a_2$ and $b_2$. These are all $\pm 1$. If the spin components are local objective properties, then for each experimental run it must hold that
\begin{equation}
a_1(a_2 + b_2) + b_1(a_2 - b_2) = \pm2.
\end{equation}
It follows that the correlation functions, defined as  the expectation value of the product of the measurement outcomes, $E(a,b) = \overline{ab}$, must satisfy the inequality
\begin{equation}
|E(a_1,a_2)+E(a_1,b_2)+E(b_1,a_2)-E(b_1,b_2)| \leq 2
\end{equation}
for any measurement directions ${\bf a}_1, {\bf b}_1, {\bf a}_2$ or ${\bf b}_2$. In fact, one realizes after some thought that the  inequality
\begin{equation}
|E(a_1,a_2)+E(a_1,b_2)|+|E(b_1,a_2)-E(b_1,b_2)| \leq 2
\end{equation}
must be satisfied. If either $E(a_1,a_2)+E(a_1,b_2)$ or $E(b_1,a_2)-E(b_1,b_2)$ is negative, we can always reverse the measurement direction ${\bf a}_1$ or ${\bf b}_1$ so that the terms become both positive or both negative. Therefore the latter form of the inequality must also hold.

For the quantum mechanical singlet state,
\begin{equation}
|\psi^-\rangle = \frac{1}{\sqrt{2}}\left(|+\rangle_1|-\rangle_2 -
|-\rangle_1|+\rangle_2\right),
\end{equation}
the CHSH inequality is violated for some choices of measurement directions. If, however, we require that a joint quantum mechanical measurement of spin along both directions ${\bf a}_1$ and ${\bf b}_1$ is made on the first spin 1/2 particle, then the CHSH inequality will always be satisfied for the resulting measurement results \cite{jointspin, demuynck}. In this case, there are simultaneous results $a_1$ and $b_1$, as well as either $a_2$ or $b_2$, for each experimental run.

For the CHSH setup, Fine has shown that the following statements are equivalent \cite{fine}: (i) Bell inequalities are satisfied. (ii) A description using hidden variables is possible. (iii) There exist joint probability distributions for all triples of observables. The equivalence of these statements also implies
that if joint probability distributions exist for all triples of observables, then a (classical) joint probability distribution necessarily exists for all four observables. This is the reason why we only need to assume a joint measurement of spin on one of the particles, not both, for the CHSH inequality to be satisfied. 

\section{Mermin's inequality for GHZ states}
\label{ghzsec}
Suppose that we have three spin 1/2 quantum systems and that we measure either $\hat{\sigma}_x$ or
$\hat{\sigma}_y$ on each of these. If the spin components are local objective properties, then expressions involving their values should be well defined, even if only one spin component for each particle is actually measured in each run of the experiment. Let us denote the values of the spin components for particle $i$ with $x_i$ and $y_i$. It is easily verified that for each experimental run we must have \cite{mermin}
\begin{equation}
-2\leq x_1 x_2 x_3 - x_1 y_2 y_3 - y_1 x_2 y_3 - y_1 y_2 x_3 \leq 2
\label{GHZineq}
\end{equation}
because the spin components $x_i$ and $y_i$ can take only the values $\pm1$. In fact one realizes that a similar inequality would hold for any measurement directions for each of the particles. The reason for choosing $x$ and $y$ is, that for the quantum GHZ state,
\begin{equation}
|GHZ\rangle =
\frac{1}{\sqrt{2}}\left(|+\rangle_1|+\rangle_2|+\rangle_3 +
|-\rangle_1|-\rangle_2|-\rangle_3\right),
\end{equation}
we find that
\begin{equation}
\label{value}
\langle
GHZ|\hat{\sigma}_{x}^1\hat{\sigma}_{x}^2\hat{\sigma}_{x}^3-
\hat{\sigma}_{x}^1\hat{\sigma}_{y}^2\hat{\sigma}_{y}^3-
\hat{\sigma}_{y}^1\hat{\sigma}_{x}^2\hat{\sigma}_{y}^3 -
\hat{\sigma}_{y}^1\hat{\sigma}_{y}^2\hat{\sigma}_{x}^3
|GHZ\rangle = 4
\end{equation} 
because the GHZ state is an eigenstate of all the four operator combinations $\hat{\sigma}_{x}^1\hat{\sigma}_{x}^2\hat{\sigma}_{x}^3$,
$\hat{\sigma}_{x}^1\hat{\sigma}_{y}^2\hat{\sigma}_{y}^3$,
$\hat{\sigma}_{y}^1\hat{\sigma}_{x}^2\hat{\sigma}_{y}^3$ and 
$\hat{\sigma}_{y}^1\hat{\sigma}_{y}^2\hat{\sigma}_{x}^3$ in Eq.(\ref{value}). The GHZ state maximally violates inequality (\ref{GHZineq}).

\subsection{Mermin's inequality and joint quantum measurements}

In a joint measurement, whether quantum mechanical or not, there will be results $\pm 1$ for the measured spin components. Therefore inequality (\ref{GHZineq}) must be satisfied, at least if we demand that both $x$ and $y$ components are measured on each of the three systems. This means that Mermin's inequality being satisfied is a necessary condition for a joint quantum mechanical measurement to exist. It does not, however, have to be a sufficient condition.

In fact, in order for a joint quantum measurement to satisfy inequality (\ref{GHZineq}), it suffices to require that the $x$ and $y$ components are measured on systems 1 and 2 only.
In order to see this, we start with the condition for the joint measurements on one of two spin components given in Eq. (\ref{conjoint}). Condition (\ref{conjoint}) means that unless the measurement directions $\bf a$ and $\bf b$ are parallel, both $\alpha$ and $\beta$ have to be less than one. Now, with $\bf a = x$ and $\bf b=y$, we find using condition (\ref{conjoint}) that $\alpha^2 +\beta^2 \leq 1$ must hold. In terms of the three-particle correlation function $E(x_1,x_2,x_3) = \overline{x_1 x_2 x_3}$ and other similar combinations, inequality (\ref{GHZineq}) reads
\begin{equation}
|E(x_1, x_2, x_3) - E(x_1, y_2, y_3) - E(y_1, x_2, y_3) - E(y_1,
y_2, x_3)| \leq 2.
\end{equation}
The quantum-mechanical average of the correlation function ${E_{QM}}(x_1,x_2,x_3) = \langle\psi |\hat{\sigma}_{x}^1\otimes \hat{\sigma}_{x}^2\otimes \hat{\sigma}_{x}^3|\psi\rangle$, if the measurements of the spins along the $x$ directions for particles 1 and 2 are made jointly with measurements of the spins along $y$, will be
\begin{equation}
{E_{QM}}(x^J_{1},x^J_{2},x_3) = \alpha_1 \alpha_2\langle\psi
|\hat{\sigma}_{x}^1\otimes \hat{\sigma}_{x}^2\otimes
\hat{\sigma}_{x}^3|\psi\rangle ,
\end{equation}
where $J$ denotes a joint measurement. This relation is valid for any state $|\psi\rangle$, and follows from the fact that expectation values for $\hat{\sigma}_{x}^1$ and $\hat{\sigma}_{x}^2$ are scaled by the factors $\alpha_1$ and $\alpha_2$ for any measured state (see the appendix for a proof).
Similarly, we find that
\begin{eqnarray}
{E_{QM}}(y^J_{1},y^J_{2},x_3) &=& \beta_1\beta_2\langle\psi
|\hat{\sigma}_{y}^1\otimes \hat{\sigma}_{y}^2\otimes
\hat{\sigma}_{x}^3|\psi\rangle \nonumber\\
{E_{QM}}(y^J_{1},x^J_{2},y_3) &=&\beta_1\alpha_2\langle\psi
|\hat{\sigma}_{y}^1\otimes \hat{\sigma}_{x}^2\otimes
\hat{\sigma}_{y}^3|\psi\rangle \\
{E_{QM}}(x^J_{1},y^J_{2},y_3) &=& \alpha_1\beta_2\langle\psi|
\hat{\sigma}_{x}^1\otimes \hat{\sigma}_{y}^2\otimes
\hat{\sigma}_{y}^3|\psi\rangle .\nonumber
\end{eqnarray}
For the GHZ state, we therefore find that
\begin{eqnarray}
&&|{E_{QM}}(x_1^J, x_2^J, x_3) - {E_{QM}}(x_1^J, y_2^J, y_3)\nonumber\\
&&- {E_{QM}}(y_1^J, x_2^J, y_3)- {E_{QM}}(y_1^J, y_2^J, x_3)|
\nonumber\\
&&= \alpha_1\alpha_2+\alpha_1\beta_2+\beta_1\alpha_2+\beta_1\beta_2\nonumber\\
&&=(\alpha_1+\beta_1)(\alpha_2+\beta_2).
\end{eqnarray}
As $\alpha_1^2+\beta_1^2\leq 1$ and $\alpha_2^2+\beta_2^2\leq
1$, it follows that $\alpha_1+\beta_1 \leq \sqrt{2}$ and
$\alpha_2+\beta_2 \leq \sqrt{2}$, so that
\begin{eqnarray}
\label{jointGHZineq}
&&|{E_{QM}}(x^J_1, x^J_2, x_3) - {E_{QM}}(x^J_1, y^J_2, y_3)\nonumber\\
&&- {E_{QM}}(y^J_1, x^J_2, y_3) - {E_{QM}}(y^J_1, y^J_2, x_3)|
\leq 2,
\end{eqnarray}
with equality holding if and only if $\alpha_1 = \beta_1 = \alpha_2 = \beta_2 = 1/\sqrt{2}$. To summarize, requiring that quantum mechanical joint measurements of spin along $\bf x$ and $\bf y$ are made on particles 1 and 2 (as opposed to measurements of spin along {\em either} $\bf x$ {\em or} spin along $\bf y$) means that inequality (\ref{GHZineq}), or rather the average of this inequality, is {\em satisfied} for the GHZ state. Moreover, we know that there always is a joint quantum measurement which
achieves equality in condition (\ref{conjoint}). Therefore there is always a pair of joint quantum measurements for particles 1 and 2, so that the equality is reached in (\ref{GHZineq}) and (\ref{jointGHZineq}).

The fact that we only have to demand that a joint measurement is made on systems 1 and 2, and not on system 3, is similar to the situation for the CHSH inequality. For three particles with two measurement settings each, a (classical) joint probability distribution for all six observables necessarily exists if a
probability distribution exists for all quintets of observables. This follows from Fine's result relating to the CHSH situation, where the existence of joint probability distributions for all triples of observables implies the existence of a joint probability for all four observables \cite{fine}.

\subsection{Mermin-type inequality for joint measurements along any spin directions}

We can also prove that an inequality of a form similar to Mermin's inequality must be satisfied for joint quantum mechanical measurements, when the measurements are made along any spin directions. Let us assume that, on the first particle, we make a joint measurement of spins along directions ${\bf a}_1$ and ${\bf b}_1$, with results $a_1^J$ and $b_1^J$, and on the second particle a joint measurement along directions ${\bf a}_2$ and ${\bf b}_2$, with results $a_2^J$ and $b_2^J$. On the third particle we measure spin either along direction ${\bf a}_3$ or ${\bf b}_3$, with measurement outcomes $a_3$ and $b_3$. The measurement outcomes for any spin component is either +1 or -1. Clearly the product of any measurement results on the first and second particles will also take either the value +1 or -1. Assuming that the measurement
direction for particle 3 is ${\bf a}_3$, we can write
\begin{eqnarray}
\label{halfineq1}
&&p(a_1^J b_2^J=b_1^J a_2^J)\nonumber\\
&&= p(a_1^J b_2^J=b_1^J a_2^J=a_3) + p(a_1^J b_2^J=b_1^J a_2^J=-a_3) \nonumber\\
&&\geq | p(a_1^J b_2^J=b_1^J a_2^J=a_3) - p(a_1^J b_2^J=b_1^J a_2^J=-a_3)| \nonumber\\
&&= \frac{1}{2} |E(a_1^J,b_2^J,a_3) + E(b_1^J,a_2^J,a_3)|.
\end{eqnarray}
Here we have assumed that well-defined joint probabilities exist for all the jointly measured observables. This is a necessary condition for joint measurements. In a similar way, assuming that the measurement direction for particle 3 is $\bf b_3$, we can write
\begin{eqnarray}
\label{halfineq2}
&&p(a_1^Ja_2^J=-b_1^J b_2^J)\nonumber\\
&&= p(a_1^Ja_2^J=-b_1^J b_2^J=b_3) + p(a_1^Ja_2^J=-b_1^J b_2^J=-b_3)\nonumber\\
&&\geq | p(a_1^Ja_2^J=-b_1^J b_2^J=b_3) - p(a_1^Ja_2^J=-b_1^J b_2^J=-b_3)| \nonumber\\
&&= \frac{1}{2} |E(a_1^J,a_2^J,b_3) - E(b_1^J,b_2^J,b_3)|.
\end{eqnarray}
The probabilities $p(a_1^Jb_2^J=b_1^Ja_2^J)$ and $p(a_1^Ja_2^J=-b_1^Jb_2^J)$ do not depend on the direction along which spin is measured on the third particle. This follows from the no-signaling condition \cite{ghirardi}. Furthermore, we can check that 
$p(a_1^Jb_2^J=b_1^Ja_2^J)+p(a_1^Ja_2^J=-b_1^Jb_2^J)=1$, for example by noting that
$p(a_1^Jb_2^J=b_1^Ja_2^J)=p(a_1^Jb_2^Jb_1^Ja_2^J=1)$ and $p(a_1^Ja_2^J=-b_1^Jb_2^J)= p(a_1^Ja_2^Jb_1^Jb_2^J=-1)$. By adding the inequalities (\ref{halfineq1}) and (\ref{halfineq2}) we get
\begin{eqnarray}
&&|E(a_1^J,b_2^J,a_3) +
E(b_1^J,a_2^J,a_3)|\nonumber\\&&~~~~~~+|E(a_1^J,a_2^J,b_3) -
E(b_1^J,b_2^J,b_3)|\leq 2,
\end{eqnarray}
which can also be written as
\begin{eqnarray}
&&|E(a_1^J,b_2^J,a_3) +
E(b_1^J,a_2^J,a_3)\nonumber\\&&~~~~~~+E(a_1^J,a_2^J,b_3) -
E(b_1^J,b_2^J,b_3)|\leq 2.
\end{eqnarray}
As for the CHSH inequality, these two forms of the inequality are equivalent, as they have to hold for all measurement directions. This is a Mermin-type inequality for three-particle correlations. Choosing all the directions ${\bf a} _i$ to be $\bf x$ and all the  directions ${\bf b}_i$ to be  $\bf y$, we get the inequality (\ref{GHZineq}). We have shown therefore, that if we require joint probability distributions to exist for the result combinations
$(a_1b_2,b_1a_2,a_3)$ and $(a_1a_2,b_1b_2,b_3)$, which is clearly a necessary condition for a joint quantum measurement to exist, then the GHZ inequality will be {\em satisfied}. Joint quantum measurements will therefore satisfy the GHZ inequality.

\subsection{The CHSH inequality, Mermin's inequality
and joint quantum measurements}

We will now discuss a connection between the CHSH and Mermin inequalities in the context of joint  quantum measurements. We can rewrite inequality (\ref{GHZineq}) as
\begin{eqnarray}
&-2 \leq \frac{1}{2}[( x_1+y_1)( x_2 x_3 -  y_2 y_3 - x_2 y_3 -  y_2 x_3) \nonumber\\
&+( x_1-y_1)( x_2 x_3 -  y_2 y_3 + x_2 y_3 +  y_2 x_3) ] \leq 2.
\label{ghz2}
\end{eqnarray}
We have just shown, that if we make a joint quantum measurement of spin along $\bf x$ and $\bf y$ on any two of the three quantum systems, then this inequality, or more precisely its average, is satisfied.

If we make a joint quantum measurement of spin along $\bf x$ and $\bf y$ only on quantum system 1, then either $x_1+x_2$ or $x_1-x_2$ is equal to zero. The absolute value of the other nonzero quantity is equal to 2. Furthermore, we know that the maximum value of both $|E(x_2, x_3) -  E(y_2, y_3) -E( x_2, y_3) - E( y_2, x_3)|$ and $|E(x_2, x_3) -  E(y_2, y_3) +E( x_2, y_3) + E( y_2, x_3)|$ is $2\sqrt{2}$. This is Cirel'son's bound \cite{cirel}, i.e. the maximum violation of the CHSH inequality which is possible within quantum mechanics. It can be demonstrated that this bound is a consequence of quantum complementarity \cite{steveanthony}.
The average of the expression in the middle of equation (\ref{ghz2}) will therefore lie between $-2\sqrt{2}$ and $+2\sqrt{2}$. Also if we make a joint measurement only on particle 2 or 3 instead of particle 1, the average will lie between $-2\sqrt{2}$ and $+2\sqrt{2}$.

Finally, if we do not require any joint quantum measurements at all, we know that for the GHZ state the middle expression in inequality (\ref{ghz2}) takes the value 4. We thus see that requiring joint measurements on two, one, or none of the three quantum systems means that the maximum absolute value of the average of the expression in inequality
(\ref{ghz2}) goes from 2, which sets the local realistic bound, to
$2\sqrt{2}$, and to to 4, respectively \cite{steveanthony2}.

\section{Gisin's inequality with three measurement settings}
\label{gisinsec}

We proceed to discuss some cases where the restrictions for quantum mechanical joint measurements to exist are not equivalent to the corresponding local realistic bounds on correlation functions. Gisin has considered generalizations of Bell inequalities for two spin 1/2 particles and $N$ measurement settings for each particle \cite{gisinbell}. For three measurement directions ${\bf a}_1, {\bf b}_1$ and ${\bf c}_1$ for the first particle, and ${\bf a}_2, {\bf b}_2$ and ${\bf c}_2$ for the second, the inequality
\begin{equation}
\label{gisinineq}
a_1(a_2+b_2+c_2)+b_1(a_2+b_2-c_2)+c_1(a_2-b_2-c_2) \leq 5
\end{equation}
has to hold as all $a_i$, $b_i$ and $c_i$ can be assigned values +1 or -1. One now chooses co-planar directions ${{\bf a}_2} = (1,0,0)$, ${\bf b}_2 = (\cos(\pi/3),\sin(\pi/3),0)$, and ${\bf c}_2 = (\cos(2\pi/3),\sin(2\pi/3),0)$, with ${\bf a}_1$ antiparallel to ${\bf a}_2+b_2+c_2$, $\bf b_1$ antiparallel to ${\bf a}_2+{\bf b}_2-{\bf c}_2$, and ${\bf c}_1$ antiparallel to ${\bf a}_2-{\bf b}_2-{\bf c}_2$.  For the singlet state, the correlation functions are given by $E(a,b) = -\bf a\cdot b$, and one therefore obtains
\begin{eqnarray}
&&~~E(a_1,a_2)+E(a_1,b_2)+E(a_1,c_2)\nonumber\\
&& +E(b_1,a_2)+E(b_1,b_2)-E(b_1,c_2)\\
&&+ E(c_1,a_2)-E(c_1,b_2)-E(c_1,c_2)\nonumber\\
&&=|{{\bf a}_2+{\bf b}_2+ {\bf c}_2}| +|{{\bf a}_2+{\bf b}_2- {\bf c}_2}| +|{{\bf a}_2-{\bf b}_2-{\bf c}_2}|\nonumber\\
&&=6\nonumber,
\end{eqnarray}
which is clearly greater than 5. The left hand side (LHS) of the previous inequality is just the LHS of inequality (\ref{gisinineq}), rewritten in terms of correlation functions. The violation ratio is therefore 6/5 for the singlet state.

One might ask why we did not instead consider the inequality
\begin{eqnarray}
&&a_1(a_2+b_2+c_2)+b_1(a_2+b_2-c_2)\nonumber\\
&&+c_1(a_2-b_2-c_2) +d_1(a_2-b_2+c_2)\leq 6, \label{fourbell}
\end{eqnarray}
which includes a fourth combination $a_2-b_2+c_2$, absent from inequality  (\ref{gisinineq}). The answer is, that for the quantum singlet state and the co-planar directions we have chosen, ${\bf a}_2-{\bf b}_2+{\bf c}_2$ would be zero.  This term would therefore not contribute to the expression involving the correlation functions. As a consequence, inequality (\ref{fourbell}) is not violated by quantum mechanics. It is more advantageous to only include three co-planar vectors ${\bf a}_2\pm {\bf b}_2 \pm {\bf c}_2$, choosing ${{\bf a}_2,
{\bf b}_2, {\bf c}_2}$ so that the length of the fourth vector is zero.

We could also choose ${\bf a}_2={\bf x}, {\bf b}_2={\bf y}$ and ${\bf c}_2={\bf z}$, so that the length of all  four vectors ${\bf a}_2\pm {\bf b}_2 \pm {\bf c}_2$ is $\sqrt{3}$. We would then choose four vectors ${\bf a}_1, {\bf b}_1, {\bf c}_1$ and ${\bf d}_1$ antiparallel to these vectors, so that the correlations are maximal for the singlet state. But even in this case the violation of the resulting Bell inequality would be less than for inequality (\ref{gisinineq}) and co-planar directions. The expression of the correlation functions would have the maximal value $4\sqrt{3}$, and the violation ratio would be $4\sqrt{3}/6\approx 1.155$, which is less than $6/5=1.2$. The violation is bigger for co-planar vectors as the length of the sums and differences of three co-planar vectors can be made larger than for the choice ${\bf a}_2={\bf x}, {\bf b}_2={\bf y}$ and ${\bf c}_2={\bf z}$.

Let us now assume that we make joint quantum mechanical measurements of spin in the three directions ${\bf a}_2,{\bf b}_2$ and ${\bf c}_2$. Strictly speaking we only know what the restriction on a joint measurement of spin along two directions is, not what the restriction is for three directions. For the co-planar choice of ${\bf a}_2 = (1,0,0)$, ${\bf b}_2 = (\cos(\pi/3),\sin(\pi/3),0)$, and ${\bf c}_2 =
(\cos(2\pi/3),\sin(2\pi/3),0)$, however, it turns out that a joint measurement of spin in the directions ${\bf a}_2$ and ${\bf c}_2$ allows us to infer a result for direction ${\bf b}_2$, as this is a linear combination of the two other directions. The condition for a joint measurement along directions ${\bf a}_2$ and ${\bf c}_2$ is
\begin{equation}
|\alpha {\bf a}_2+\gamma{\bf c}_2| +|\alpha {\bf} a_2-\gamma {\bf
c}_2| \leq 2.
\end{equation}
If we take $\alpha=\gamma$, this condition means that $\alpha=\gamma\leq 2/(1+\sqrt{3})$. The jointly measured expectation values we would arrive at are
\begin{eqnarray}
\overline{a_2} &=& \alpha\langle {\bf a}_2\cdot
\hat{\sigma}\rangle, \quad  \overline{c_2} = \alpha\langle {\bf
c}_2\cdot
\hat{\sigma}\rangle \quad \text{and}\nonumber \\
\quad \overline{b_2} &=& \alpha\langle ({\bf a}_2 + {\bf c}_2) \cdot
\hat{\sigma}\rangle = \alpha\langle {\bf b}_2\cdot
\hat{\sigma}\rangle .
\end{eqnarray}
The correlation functions for the joint measurement will again be proportional to those of the corresponding sharp measurements, $E^J(a_1,a_2) = \alpha E(a_1,a_2)$, and similarly for other pairs of directions. We therefore get
\begin{eqnarray}
&&~~E^J(a_1,a_2)+E^J(a_1,b_2)+E^J(a_1,c_2)\nonumber\\
&&+E^J(b_1,a_2)+E^J(b_1,b_2)-E^J(b_1,c_2)\\
&&+ E^J(c_1,a_2)-E^J(c_1,b_2)-E^J(c_1,c_2)\nonumber \\
&&=\alpha\left(|{\bf a}_2+{\bf b}_2+ {\bf c}_2| +|{\bf a}_2+{\bf b}_2- {\bf c}_2| +|{\bf
a}_2-{\bf b}_2- {\bf c}_2|\right)\nonumber
\\&&\leq \frac{2}{1+\sqrt{3}}6 \approx 4.392 < 5.\nonumber
\end{eqnarray}
If we require making joint measurements of the three co-planar directions $\bf a_2,b_2$ and $\bf c_2$, then the Bell inequality given in (\ref{gisinineq}) is clearly not violated. Furthermore, it is not possible to reach equality in (\ref{gisinineq}). This is in contrast to the case for the CHSH and GHZ inequalities, where it was possible at least to reach equality when making joint quantum mechanical measurements on all but one of the quantum systems.

What happens if we consider joint measurements along three mutually orthogonal directions? The condition restricting joint measurements of spin along three directions ${\bf a}, {\bf b}$ and ${\bf c}$ should be most restrictive when the three directions are as incompatible as possible. It is easy to show (see \cite{busch}) that a sufficient condition for a joint measurement along three directions is
\begin{equation}
\label{threecond}
\alpha^2+\beta^2+\gamma^2 \leq 1,
\end{equation}
where $\alpha, \beta, \gamma$ are the factors by which the expectation values for spin along $\bf a,b$ and ${\bf c}$ must scale down. This condition is clearly not always necessary for arbitrary measurement directions. It is evidently too restrictive for example when the directions are co-planar, or when two of the
directions are the same. We believe, however, that it is indeed necessary when {\bf a} = {\bf x}, {\bf b} = {\bf y} and {\bf c} = {\bf z}. In this case, it would follow from condition (\ref{threecond}) that
\begin{eqnarray}
\alpha \left( |{\bf x+y+ z}| +|{\bf x+y- z}| ~\right. & \\
\left. +|{\bf x-y- z}|+|{\bf x-y+ z}|\right) & \leq 4 < 6,
\nonumber
\end{eqnarray}
when $\alpha=\beta=\gamma$. Making a joint measurement along the ${\bf x,y}$ and $\bf z$ directions would force the correlations to keep well below the local realistic bound.

\section{Conclusions}
\label{conclsec} 
We have investigated the similarities and differences between Bell inequalities and the conditions for
joint quantum mechanical measurements. In a Bell inequality, one considers measurements on two or more particles. On each particle, one can choose between two or more measurement directions. Local realism, or the assumption that we can assign a value to each quantity, whether measured or not, leads us to inequalities for the correlation functions of the measurements. Some quantum mechanical states are found to violate these inequalities, if we only measure one of the involved
observables on each quantum system in each run of the experiment.

If, however, we make joint quantum mechanical measurements of all the observables involved, then the corresponding Bell inequality always has to be satisfied. This follows from the fact that performing a joint measurement means that values for each of the observables necessarily exist in each run of the experiment. The CHSH inequality and Mermin's inequality for GHZ states turn out to be equivalent to the restrictions posed by making joint quantum measurements. Bell inequalities, however, are not always sufficient conditions for joint quantum measurements to exist. We have shown that Gisin's inequality for three co-planar directions is found to be weaker than the condition placed by making joint measurements of the involved observables.

It remains to be understood exactly why the joint measurement condition and the CHSH and Mermin inequalities are equivalent.

\acknowledgments
W. Son and M.S. Kim want to thank the UK Engineering and Physical Science Research Council and KRF (2003-070-C00024) for financial support. E. Andersson gratefully acknowledges financial support from the the Dorothy Hodgkin Fellowship scheme of the Royal Society of London.

\appendix*
\section{Correlation functions for joint measurements}

For a joint measurement of the spin observables $\hat{A}=\bf a\cdot\vec\sigma$ and $\hat{B}=\bf b\cdot\vec\sigma$, we have required that
\begin{equation}
\label{scalingeq}
\overline{A_J} = \alpha\langle \hat{A}\rangle ~~~\mbox{and}~~~
\overline{B_J} = \beta\langle\hat{B}\rangle
\end{equation}
hold for any measured state. Here $\alpha$ and $\beta$ are positive constants and $\overline{A_J}$ and $\overline{B_J}$ are expectation values of the jointly measured observables. In this appendix, we will show that correlation functions scale in the same way as the eigenvalues.

Suppose that we have two spin 1/2 particles in a state $|\psi_{12}\rangle$, which may be entangled. For simplicity, we here present the case of a pure state, but the proof can easily be generalized to mixed states. We are making a joint measurement of spin along $\bf a$ and $\bf b$ on system 1. On the second system, we measure spin along direction $\bf c$. Let us consider the correlation for the measurements of spin along $\bf a$ and $\bf c$. Denoting the measurement results by $a$ and $c$, the correlation function is given by
\begin{eqnarray}
E_J(a,c)= [p(a=+|c=+)-p(a=-|c=+)]p(c=+)\nonumber\\
-[p(a=+|c=-)-p(a=-|c=-)]p(c=-)\nonumber
\end{eqnarray}
where $p(a=x|c=y)$ means the conditional probability of obtaining $a=x$ given that $c=y$ has been obtained, and $J$ denotes that a joint measurement was made on system 1.
Let us denote the conditional state of system 1, given that $c=+$ was obtained, by $|\psi_1(c=+)\rangle$, and the conditional state, given that $c=-$ was obtained, by $|\psi_1(c=-)\rangle$. For both these states, as for any state of system 1, we know that the expectation value for the jointly measured spin observable changes according to equation (\ref{scalingeq}). Moreover, the conditional states are independent of whether or not a joint measurement is made on system 1. Therefore we have that
\begin{eqnarray}
E_J(a,c) &=& \alpha\langle\psi_1(c=+)|{\bf a}\cdot\vec\sigma|\psi_1(c=+)\rangle p(c=+)\nonumber\\
&&-\alpha\langle\psi_1(c=-)|{\bf a}\cdot\vec\sigma|\psi_1(c=-)\rangle p(c=-)\nonumber\\
&=& \alpha E(a,c),
\end{eqnarray}
where $E(a,c)$ is the correlation function in the case where the spin observables are measured on their own on both systems 1 and 2.


\begin{thebibliography}{00}
\bibitem{artkel} E. Arthurs and J. l. Kelly, Bell Syst. Tech. {\bf 44}, 725 (1965).
\bibitem{artgood} E. Arthurs and M. S. Goodman, Phys. Rev. Lett. {\bf 60}, 2447 (1988).
\bibitem{stig} S. Stenholm, Ann. Phys. N.Y. {\bf 218}, 233 (1992).
\bibitem{walker} N. G. Walker and J. E. Carroll, Electron. Lett. {\bf 20}, 981 (1984); N. G. Walker, J. Mod. Opt. {\bf 34}, 15 (1987).
\bibitem{barnett} S. M. Barnett and P. M. Radmore, Methods in Theoretical Quantum Optics, Oxford, 1997.
\bibitem{bell} J. S. Bell, {\it Speakable and unspeakable in quantum mechanics}, Cambridge University Press, Cambridge, 1987; new edition 2004.
\bibitem{fine} A. Fine, Phys. Rev. {\bf 48}, 291 (1982).
\bibitem{jointspin} E. Andersson, S. M. Barnett and A. Aspect, submitted to Phys. Rev. A (2005).
\bibitem{demuynck} W. De Muynck and O. Abu-Zeid, Phys. Lett. {\bf 100A}, 485 (1984).
\bibitem{chsh} J. F. Clauser, M. A. Horne, A. Shimony and R. A. Holt, Phys. Rev. Lett. {\bf 23}, 880 (1969).
\bibitem{mermin} N. D. Mermin, Phys. Rev. Lett. {\bf 65}, 1838 (1990).
\bibitem{greenberger} D. M. Greenberger, M. A. Horne, and A. Zeilinger,   in Bell's Theorem, Quantum Theory, and Conceptions of the Universe, edited by M. Kafatos (Kluwer, Dordrecht, 1989); D. M. Greenberger, M. A. Horne, A. Shimony, and A. Zeilinger, Am. J. Phys. 58, 1131 (1990).
\bibitem{gisinbell} N. Gisin, Phys. Lett. A {\bf 260}, 1-3 (1999).
\bibitem{busch} P. Busch, Phys. Rev. D {\bf 33}, 2253 (1986); P. Busch, M. Grabowski, and P. J. Lahti,
Operational quantum physics, Springer-Verlag, Berlin, 1995, p. 109-110.
\bibitem{hel} C. W. Helstrom, Quantum Detection and Estimation Theory, Academic Press, New York, 1976.
\bibitem{nch} M. A. Nielsen and I. L. Chuang, Quantum Computation and Quantum Information,
Cambridge University Press, Cambridge, 2000.
\bibitem{ghirardi} G. C. Ghirardi, A. Rimini and T. Weber, Lett. Nuovo Cim. {\bf 27}, 293 (1980).
\bibitem{cirel} B. S. Cirel'son, Lett. Math. Phys. {\bf 4}, 93 (1980).
\bibitem{steveanthony} A. Chefles and S. M. Barnett, J. Phys. A: Math. Gen. {\bf 29}, L237 (1996).
\bibitem{steveanthony2} A. Chefles and S. M. Barnett, Phys. Rev. A {\bf 55}, 1721 (1997).
\bibitem{buschbeam} P. Busch, Found. Phys. {\bf 17}, 905 (1987).
\bibitem{bar} S. M. Barnett and E. Andersson, Phys. Rev. A {\bf 65}, 044307 (2002).
\bibitem{gisin} N. Gisin, Phys. Lett. A {\bf 242}, 1 (1998).
\bibitem{dagmar} D. Bruss, G. M. D'Ariano, C. Macchiavello, and M. F. Sacchi, Phys. Rev. A {\bf 62}, 062302 (2000).
\bibitem{rob2} H. P. Robertson, Phys. Rev. {\bf 46}, 794 (1934).
\end{thebibliography}
\end{document}